\documentclass[a4paper]{article}
\usepackage[affil-it]{authblk}
\usepackage[usenames,dvipsnames]{xcolor}
\usepackage{amsfonts}
\usepackage{amsmath,amsthm,amssymb,dsfont,pifont}
\usepackage{enumerate}
\usepackage[english]{babel}
\usepackage{graphicx}	
\usepackage{subcaption}
\usepackage[margin=3cm]{geometry}
\usepackage{url}
\usepackage{todonotes}
\usepackage{bbm}
\usepackage{caption}
\usepackage{subcaption}
\usepackage{tikz}
\usetikzlibrary{chains}
\usetikzlibrary{fit}
\usetikzlibrary{quantikz}
\usepackage{makecell}
\usepackage{cite}

\usepackage{epsfig}
\usetikzlibrary{shapes.symbols,patterns} 
\usepackage{pgfplots}

\usepackage{hyperref}[breaklinks]
\hypersetup{colorlinks=true,citecolor=blue,linkcolor=blue,filecolor=blue,urlcolor=blue}

\usepackage{nicefrac}
\usepackage{mathtools}

\usepackage{algorithm}
\usepackage{algorithmic}

\usepackage{optidef}

\usepackage{bold-extra}

\tikzset{meter/.append style={draw, inner sep=8, rectangle, font=\vphantom{A}, minimum width=30, line width=.8,
 path picture={\draw[black] ([shift={(.1,.3)}]path picture bounding box.south west) to[bend left=50] ([shift={(-.1,.3)}]path picture bounding box.south east);\draw[black,-latex] ([shift={(0,.1)}]path picture bounding box.south) -- ([shift={(.3,-.1)}]path picture bounding box.north);}}}
 
\theoremstyle{plain}

\theoremstyle{definition}

\newcommand*{\cE}{\mathcal{E}}
\newcommand*{\cF}{\mathcal{F}}

\newcommand*{\cO}{\mathcal{O}}

\newcommand*{\cS}{\mathcal{S}}

\newcommand*{\N}{\mathbb{N}}

\newcommand*{\R}{\mathbb{R}}

\newcommand{\norm}[1]{\left\lVert#1\right\rVert}

\definecolor{mylightgreen}{RGB}{219,255,192}
\definecolor{mylightblue}{RGB}{240,255,252}
\definecolor{mylightyellow}{RGB}{255,248,180}
\definecolor{mylightorange}{RGB}{252,248,236}
\definecolor{mygraywhite}{RGB}{243, 243,243}
\definecolor{mylightred}{RGB}{255, 209,192}
\definecolor{mylightpink}{RGB}{255,240,252}

\usepackage{float}
\DeclareMathOperator{\sign}{sign}
\newcommand{\Pegasos}{Pegasos}
\usepackage{physics}
\usepackage{algorithmic}
\usepackage{algorithm}
\usepackage{cleveref}

\usepackage{tikz}
\usetikzlibrary{quantikz}
\usetikzlibrary{shapes.symbols,patterns} 
\usepackage{pgfplots}
\usetikzlibrary{shapes.symbols,patterns} 

\definecolor{myBlue}{RGB}{31,119,180}
\definecolor{myOrange}{RGB}{255, 127, 14}
\definecolor{myGreen}{RGB}{44,160,44}
\definecolor{myRed}{RGB}{214,39,40}
\definecolor{myViolet}{RGB}{148,103,189}
\definecolor{myGrey}{RGB}{127,127,127}
\definecolor{myCyan}{RGB}{23, 190, 207}

\definecolor{featureMap}{RGB}{23, 190, 207}
\definecolor{variationalForm}{RGB}{148,103,189}
\definecolor{highlight}{RGB}{214,39,40}

\allowdisplaybreaks
\crefname{equation}{}{}
\title{Quantum Kernel Alignment with Stochastic\\  Gradient Descent}

 \author{\normalsize Gian Gentinetta$^{1}$, David Sutter$^{2}$, Christa Zoufal$^{2}$, Bryce Fuller$^{3}$, and Stefan Woerner$^{2}$}
  \affil{\small $^{1}$Institute of Physics, \'Ecole Polytechnique F\'ed\'erale de Lausanne\\
  $^{2}$IBM Quantum, IBM Research Europe -- Zurich\\
  $^{3}$IBM Quantum, IBM T.J.~Watson Research Center, Yorktown Heights
 }
 
 \date{}

\begin{document}
\maketitle

\begin{abstract}
Quantum support vector machines have the potential to achieve a quantum speedup for solving certain machine learning problems. The key challenge for doing so is finding good quantum kernels for a given data set --- a task called kernel alignment. In this paper we study this problem using the Pegasos algorithm, which is an algorithm that uses stochastic gradient descent to solve the support vector machine optimization problem. We extend Pegasos to the quantum case and and demonstrate its effectiveness for kernel alignment. Unlike previous work which performs kernel alignment by training a QSVM within an outer optimization loop, we show that using Pegasos it is possible to simultaneously train the support vector machine and align the kernel. Our experiments show that this approach is capable of aligning quantum feature maps with high accuracy, and outperforms existing quantum kernel alignment techniques. Specifically, we demonstrate that Pegasos is particularly effective for non-stationary data, which is an important challenge in real-world applications.
\end{abstract}

\maketitle

\section{Introduction}
Finding a practically relevant problem that can be solved better or faster with a quantum computer compared to the best classical implementation is a grand challenge in the field of quantum computing. Quantum machine learning problems are potential candidates for demonstrating a quantum advantage~\cite{QML_lloyd17,Havlicek2019,Abbas2020a,temme21} and it has been shown for certain artificially constructed data sets that quantum support vector machines (QSVMs) offer an exponential speedup compared to any known classical algorithm~\cite{temme21}. It is an ongoing topic of research to determine whether similar speedups are heuristically available for practical problems. 

The performance of QSVMs are dependent upon the degree to which a quantum kernel encodes information about the problem~\cite{NEURIPS2021_69adc1e1}, yet little is known about how to engineer quantum kernels which exploit the structure of an arbitrary dataset. One solution to this problem is to choose a parameterized family of quantum kernel functions and attempt to learn a kernel which is well suited to a given dataset. This process of localizing a good kernel for class of given data is referred to as \emph{kernel alignment}~\cite{cristiani01,cortes12,Glick2021}. A core limitation of this approach is that apriori it is not easy to optimize over a parameterized space of kernels and the overheads introduced can be prohibitive.

In this work, we present a novel method to speedup the kernel alignment problem via the use of the Pegasos algorithm~\cite{Shalev-Shwartz2008,ComplexityOfQSVM}.
Pegasos is an iterative gradient-based algorithm that solves the primal SVM optimization problem and allows for a direct integration of kernel alignment. Trainable parameters in the quantum circuit that implements a feature map can be optimized simultaneously with the weights of the SVM problem. This drastically reduces the complexity of quantum kernel alignment (QKA) compared to the dual approach~\cite{Glick2021} where the full SVM problem has to be solved as a subroutine in the optimization of the kernel parameters. 
Furthermore, the iterative nature of Pegasos is ideal for online machine learning and non-stationary data as it enables easy continuation of training as well as unlearning the impact of outdated training data.

In an experimental demonstration we show that we can successfully classify the covariant data set proposed in~\cite{Glick2021} using Pegasos kernel alignment. We further show in a hardware experiment that Pegasos can adapt to non-stationary data in real-time by continuously retraining the parametrization while keeping the training accuracy high.
This may be relevant in practice where non-stationarity is unavoidable.
\section{Support vector machines}
\subsection{Classical support vector machines}
Given an unknown probability distribution $P(\mathbf{x},y)$ for data vectors $\mathbf{x} \in \R^r$ and class membership labels $y \in \{-1,1\}$, we draw a set of training data $X = \{\mathbf{x}_1,\dots,\mathbf{x}_M\}$ with corresponding labels $y = \{y_1,\dots,y_M\}$. The support vector machine (SVM)~\cite{Vapnik1992, Vapnik2000,Cortes95} defines a classification function $c: \R^r \to \{-1,1\}$ implementing the trade-off between accurately predicting the true labels and maximizing the orthogonal distance between the two classes within the training data.

For a given feature map $\varphi: \R^r \to \R^s$, the decision boundary is
\begin{equation*}
    c_{\textnormal{SVM}}(\mathbf{x}) = \sign\big[\mathbf{w}^\top\varphi(\mathbf{x})\big] \, ,
\end{equation*}
for $\mathbf{w} \in \R^s$.
L et $\mathbf{w}^\star$ denote the hyperplane defined as the solution to the primal optimization problem 
\begin{align}\label{opt:svm_soft_margin_primal_hinge_loss}
 \textnormal{(primal problem)} \qquad			\min_{\mathbf{w} \in \R^s} \left \lbrace \frac{\lambda}{2} \norm{\mathbf{w}}^2 +  \sum_{i=1}^M \max \left\lbrace 0, 1 - y_i \big\langle\mathbf{w}^,\varphi(\mathbf{x}_i)\big\rangle \right\rbrace \right \rbrace \, ,
\end{align}
where the term containing $\lambda>0$ provides regularization and the second term is a sum over the hinge losses of the data points from the training set.

The optimization problem~\cref{opt:svm_soft_margin_primal_hinge_loss} features a dual formulation as
\begin{align} \label{opt:svm_soft_margin_feature_map_dual}
\textnormal{(dual problem)} \qquad  \left \lbrace
\begin{array}{r l}
\max \limits_{\alpha_i \in \R} & \sum_{i=1}^M \alpha_i - \frac{1}{2} \sum_{i,j=1}^M \alpha_i \alpha_j y_i y_j \, k(\mathbf{x}_i, \mathbf{x}_j) - \frac{\lambda}{2} \sum_{i=1}^M \alpha_i^2 \\
\textnormal{s.t.} & 0 \leq \alpha_i \quad \forall i=1,\ldots,M \, ,
\end{array}
\right.
\end{align}
where $k(\mathbf{x},\mathbf{y}) := \langle\varphi(\mathbf{x}),\varphi(\mathbf{y})\rangle$ denotes the kernel function.
From~\Cref{opt:svm_soft_margin_feature_map_dual}, we see that solving the dual requires the evaluation of the full kernel matrix $K \in \R^{M \times M}$ defined via its entries 
\begin{equation*}
	K_{ij} = k(\mathbf{x}_i, \mathbf{x}_j) \quad \textnormal{for} \quad i,j=1,\ldots,M \, .
\end{equation*}
Given $K$, the dual optimization can be restated as a convex quadratic program (see~\cite{ComplexityOfQSVM}) and therefore be solved in polynomial time regardless of the dimension of the feature vectors ~\cite{Boyd2004}. For this reason the dual formulation is often favored in practice.

Interestingly, there is an algorithm to solve the primal problem called \Pegasos~\cite{Shalev-Shwartz2011a}, which employs stochastic sub-gradient descent to minimize the objective. Unlike a direct implementation of~\Cref{opt:svm_soft_margin_primal_hinge_loss}, \Pegasos\ only requires access to the kernel matrix $K$ and thus avoids explicitly representing high dimensional feature vectors. A description of the original algorithm and its quantum version can be found in~\cite{Shalev-Shwartz2011a,ComplexityOfQSVM}.

\subsection{Quantum support vector machines}
A quantum support vector machine is simply a support vector machine equipped with a quantum kernel function.
Because there are quantum circuits that can be run efficiently on quantum hardware but cannot be simulated efficiently on any classical computer, we can define kernel functions that can only be evaluated efficiently on a quantum computer.
Note that this does not yet answer the question if these kernels are useful.
Furthermore, while QSVMs show potential for practical quantum advantage~\cite{temme21,Glick2021} they also suffer from disadvantages such as exponential concentration phenomena inducing flat training landscapes (also called barren plateaus)~\cite{ThanasilpExponentialConcentration22, Cerezo_2021_costfunct, Wang_21_noiseinducedBPs, Wiebe2020Barren}.
In fact, exponentially vanishing gradients occur for quantum kernels \cite{ThanasilpExponentialConcentration22} if (i) the quantum feature map forms (an approximate) $2-$design; (ii) the chosen quantum feature map leads to product states whose fidelity is on average exponentially small; (iii) the quantum hardware used to evaluate the quantum kernel entries is impacted by too much noise, e.g., Pauli noise acting before and after each gate.
We can, therefore, conclude that quantum kernels have to be designed very carefully to allow for efficient trainability.

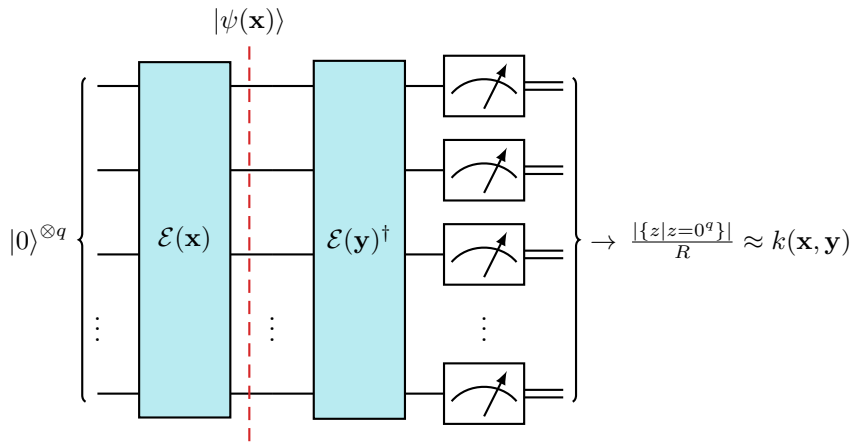
\begin{figure}[!htb]
            \centering
            \begin{quantikz}[row sep=0.2cm]
                \lstick[wires=5]{$\ket{0}^{\otimes q}$} 
                & \gate[5, nwires={4}, style={fill=myCyan!30}][1.2cm]{\cE(\mathbf{x})} \slice[style=myRed]{$\ket{\psi(\mathbf{x})}$} & \qw & \gate[5, nwires={4}, style={fill=myCyan!30}][1.2cm]{\cE(\mathbf{y})^\dagger} & \meter{} & \cw \rstick[wires=5]{$\rightarrow \, \frac{|\{ z | z = 0^q\} |}{R} \approx k(\mathbf{x}, \mathbf{y})$}\\
                & & \qw & & \meter{} & \cw\\
                & & \qw & & \meter{} & \cw\\
                \vdots & & \vdots & & \vdots &\\
                & & \qw & & \meter{} & \cw
            \end{quantikz}
            \caption{\textbf{Quantum kernel evaluation~\textnormal{\cite{Liu2021,ComplexityOfQSVM}}:} Let $\cE(\mathbf{x})$ denote a parametrized unitary which defines the feature map $\ket{\psi(\mathbf{x})}=\cE(\mathbf{x}) \ket{0}^{\otimes q}$. Preparing the state $\cE(\mathbf{y})^\dagger \cE(\mathbf{x})\ket{0}^{\otimes q}$ and  measuring all of the qubits in the computational basis, a bit string $z \in \{0, 1\}^q$ is determined. Repeating this process $R$-times, the frequency of the all zero outcome approximates the kernel value $k(\mathbf{x}, \mathbf{y})$ in~\cref{eq:quantum_kernel}.}
            \label{fig:svm_quantum_kernel_circuit}
        \end{figure}

More formally, consider a feature map
\begin{equation*}
			\begin{aligned}
				\psi \colon \R^r & \to \cS(2^q)                                          \\
				\mathbf{x}       & \mapsto \ketbra{\psi(\mathbf{x})}{\psi(\mathbf{x})} \, ,
			\end{aligned}
\end{equation*}
where $\cS(2^q)$ denotes the space of density matrices on $q$ qubits~\cite{Havlicek2019}.
The kernel function is then given by
\begin{equation}\label{eq:quantum_kernel}
		k(\mathbf{x}, \mathbf{y})  = \mathrm{tr} \big[ \ketbra{\psi(\mathbf{y})}{\psi(\mathbf{y})} \, \ketbra{\psi(\mathbf{x})}{\psi(\mathbf{x})} \big]
		                            = |\!\braket{\psi(\mathbf{x})}{\psi(\mathbf{y})}\!|^2 \, .
\end{equation}
\Cref{fig:svm_quantum_kernel_circuit} explains how to define and evaluate a kernel function via a quantum circuit.


It remains unknown how to reliably find good quantum kernels for practical datasets. Kernel alignment is one method which has shown promise in tailoring quantum kernels to specific datasets. In the remainder of this work, we propose a method for training a QSVM which allows for the quantum kernel to be simultaneously aligned to the target dataset during training.

\section{Quantum kernel alignment with Pegasos} \label{sec_QKA_Pega}
\subsection{Quantum kernel alignment}
Here we discuss how trainable parameters can be used to optimize a feature map to match the training data in the QSVM problem. Choosing the quantum circuit used as a feature map in the QSVM problem is an important and non-trivial task. The choice of the feature map decides whether the data is (linearly) separable in the feature space and, thus, whether the QSVM succeeds in classifying the data. While there are some tricks, such as exploiting symmetries of the system~\cite{Meyer2022}, these usually require some knowledge about the structure of the data. Furthermore, even when it is known that data exhibits particular structure, it is often not obvious how to build a circuit which takes advantage of this knowledge. Instead, Glick \emph{et al.}~\cite{Glick2021} proposed to automate the fine-tuning of the feature map by optimizing the training loss with respect to additional trainable parameters in the circuit.

\begin{figure}[!htb]
    \centering
    \begin{quantikz}[row sep=0.2cm]
			\lstick[wires=5]{$\ket{0}^{\otimes q}$} 
            & \gate[5, nwires={4}, style={fill=myViolet!30}][1.2cm]{\cF_0(\boldsymbol{\theta})}  
			& \gate[5, nwires={4}, style={fill=myCyan!30}][1.2cm]{\cE_1(\mathbf{x})}  
            & \gate[5, nwires={4}, style={fill=myViolet!30}][1.2cm]{\cF_1(\boldsymbol{\theta})}  
            & \gate[5, nwires={4}, style={fill=myCyan!30}][1.2cm]{\cE_2(\mathbf{x})}  
            & \ldots\qw 
            & \gate[5, nwires={4}, style={fill=myCyan!30}][1.2cm]{\cE_d(\mathbf{x})}  
            & \gate[5, nwires={4}, style={fill=myViolet!30}][1.2cm]{\cF_d(\boldsymbol{\theta})} 
            &  \qw & \rstick[wires=5]{$\ket{\psi_\theta(\boldsymbol{x})}$} \\
			& &   & & & \ldots\qw &  & & \qw &\\
			& &   & & & \ldots\qw &  & & \qw & \\
			\vdots & &   & & & \vdots &  & & \vdots &\\
			& &  &   & & \ldots\qw &  & & \qw & 
		\end{quantikz}
    \caption{\textbf{Trainable feature map}: The feature map composed of unitary gates $\cE_i(\mathbf{x})$ used to upload the datum $\mathbf{x}$ is expanded by embedding unitaries $\cF_i(\mathbf{\theta})$. The parameters $\boldsymbol{\theta}$ are trained to minimize the training loss.}
    \label{fig:trainable_circuit}
\end{figure}
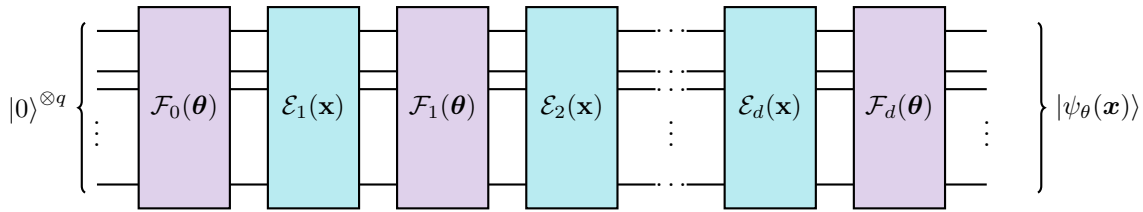

In the following, we denote the feature state prepared from our trainable quantum circuit as $\ket{\psi_\theta(\boldsymbol{x})}=\ket{\psi(\theta,\boldsymbol{x})}$. Again, the kernel function is defined as the overlap $k_\theta(\boldsymbol{x},\boldsymbol{y}) = \left| \braket{\psi_\theta(\boldsymbol{x})}{\psi_\theta(\boldsymbol{y})}\right|^2$. With this notation, we can extend the primal and dual SVM optimization problems given by~\cref{opt:svm_soft_margin_primal_hinge_loss} and~\cref{opt:svm_soft_margin_feature_map_dual}, by taking into consideration the optimization of $\theta$. This results in
\begin{align}\label{opt:primal_qka}
 \textnormal{(QKA: primal problem)} \quad \min \limits_{\theta \in \R^d} \min_{\mathbf{w} \in \R^s} \left \lbrace \frac{\lambda}{2} \norm{\mathbf{w}}^2 +  \sum_{i=1}^M \max \left\lbrace 0, 1 - y_i \big\langle\mathbf{w}, \psi_\theta(\mathbf{x}_i)\big\rangle \right\rbrace \right \rbrace \, 
\end{align}
and
\begin{equation} \label{opt:dual_qka}
\textnormal{(QKA: dual problem)} 
\left \lbrace
\begin{array}{r l}
\min \limits_{\theta \in \R^d} \max \limits_{\alpha_i \in \R} & \sum_{i=1}^M \alpha_i - \frac{1}{2} \sum_{i,j=1}^M \alpha_i \alpha_j y_i y_j \, k_\theta(\mathbf{x}_i, \mathbf{x}_j) - \frac{\lambda}{2} \sum_{i=1}^M \alpha_i^2 \\
\textnormal{s.t.} & 0 \leq \alpha_i \quad \forall i=1,\ldots,M \, .
\end{array} \right .
\end{equation}

In~\cite{Glick2021} the dual problem~\cref{opt:dual_qka} is solved through a nested optimization. In an outer loop the trainable parameters ${\theta}$ are optimized, where in every optimization step the dual QSVM problem is solved to optimize the hyperplane parameters $\boldsymbol{\alpha}$. While this is a viable option for smaller data sets, solving the dual optimization problem becomes prohibitively expensive if the size of the data set $M$ is large. In fact, to reach an accuracy of $\varepsilon$ with respect to the ideal hyperplane, a total of $\cO\left(M^{4.67}/\varepsilon^2\right)$ circuit evaluations are required~\cite{ComplexityOfQSVM}. Thus, solving the dual problem as a subroutine in the optimization of the trainable parameters makes this algorithm unfeasible for large $M$.

Instead of solving the min-max problem in \cref{opt:dual_qka}, we utilize the min-min property of the primal formulation of the QKA problem~\cref{opt:primal_qka}. 
The min-min problem has the advantage that the two minimizations can be done simultaneously where the min-max problem has to be solved sequentially which is considerably more expensive.
To solve the primal problem, the \Pegasos~algorithm is employed and adapted to also optimize the trainable parameters.

\subsection{Pegasos quantum kernel alignment}
In the following, we derive how we can adapt the stochastic gradient descent (SGD) based \Pegasos~algorithm to solve the primal formulation of the quantum kernel alignment problem~\Cref{opt:primal_qka}. \Pegasos~is a classical algorithm that finds the optimal weights $\mathbf{w}$ in the primal SVM problem through SGD~\cite{Shalev-Shwartz2008}. Compared to the more commonly used dual solvers, \Pegasos~has been shown to scale favourably for large training data sets in the presence of shot noise~\cite{ComplexityOfQSVM}.

The main idea of the \Pegasos~algorithm is to write the the weights as a linear combination of the feature vectors, i.e.,
\begin{equation*}
    \mathbf{w}_\tau = \frac{1}{\lambda\tau}\sum_{t=1}^\tau\alpha_ty_{i_t}\psi(\mathbf{x}_{i_t})\, .
\end{equation*}
Here we sum over the iterations, where $(\mathbf{x}_{i_t}, y_{i_t})$ is the datum (with features and label) sampled in iteration $t$ and $\alpha_t \in \{0,1\}$ indicates whether the datum has been chosen as a support vector. Using this ansatz for $\mathbf{w}$ allows us to write the inner product $\langle\mathbf{w},\psi(\mathbf{x}_j)\rangle$ in~\Cref{opt:primal_qka} in terms of the kernel function such that direct access to the feature vectors is not required.

We now extend this ansatz by adding trainable parameters $\theta_t$ to the feature map via
\begin{equation}
\label{eq:pegasos_it_sum}
    \mathbf{w_\tau} = \frac{1}{\lambda\tau}\sum_{t=1}^\tau\alpha_ty_{i_t}\psi_{\theta_t}(\mathbf{x}_{i_t})\, .
\end{equation}
In addition to updating $\mathbf{w}$, we also perform an update step on $\theta$. This is possible thanks to the min-min structure of \cref{opt:primal_qka}.

Assuming we know the values of $\alpha_t$ for $t<\tau$ and $\theta_t$ for $t \leq \tau$, we next derive the values for $\alpha_{\tau}$ and $\theta_{\tau+1}$ using SGD. The optimization step starts by uniformly sampling $(\mathbf{x}_{i_{\tau}}, y_{i_{\tau}})$ from the training data set. The loss function for this data point is then given as
\begin{equation}\label{eq:partial_obj}
    f^{\tau }(\theta,\mathbf{w}) = \frac{\lambda}{2}\norm{\mathbf{w}}^2 + \max\left[0,1-y_{i_{\tau}}\langle\mathbf{w},\psi_{\theta}(\mathbf{x}_{i_{\tau}})\rangle\right].
\end{equation}

We first calculate the gradient with respect to $\mathbf{w}$ as
\begin{equation*}
    \frac{\partial f^\tau}{\partial \mathbf{w}}(\mathbf{w}, \theta_\tau) =
    \begin{cases}
        \lambda \mathbf{w}, & \textnormal{if } y_{i_{\tau + 1}}\langle\mathbf{w},\psi_{\theta_\tau}(\mathbf{x}_{i_{\tau}})\rangle > 1 \\
        \lambda \mathbf{w} - y_{i_{\tau}}\psi_{\theta_\tau}(\mathbf{x}_{i_{\tau}}), & \textnormal{otherwise}\, .
    \end{cases}
\end{equation*}
The inner product in the if-condition can be evaluated using the kernel trick as
\begin{equation*}
    \langle\mathbf{w},\psi_{\theta_\tau}(\mathbf{x}_{i_{\tau}})\rangle = \frac{1}{\tau-1}\sum_{t=1}^{\tau-1}\alpha_ty_{i_t}k_{\theta_t,\theta_\tau}(\mathbf{x}_{i_t},\mathbf{x}_{i_\tau})\, ,
\end{equation*}
where we introduced the pseudo-kernel $k_{\theta, \phi}(\mathbf{x},\mathbf{y}) = \langle\psi_\theta(\mathbf{x}),\psi_\phi(\mathbf{y})\rangle = \left| \braket{\psi_\theta(\boldsymbol{x})}{\psi_\phi(\boldsymbol{y})}\right|^2$. Next, the weights are updated according to a learning rate of $\mu_{\tau} = 1/\lambda \tau$, which leads to
\begin{align*}
    \mathbf{w}_{\tau} &= \mathbf{w}_{\tau - 1} -\frac{1}{\lambda\tau}\frac{\partial f^\tau}{\partial \mathbf{w}}(\theta_\tau,\mathbf{w}_{\tau - 1}) \\ &= 
    \begin{cases}
        (1 - \frac{1}{\tau})\mathbf{w}_{\tau-1}, & \textnormal{if } y_{i_{\tau}}\langle\mathbf{w},\psi_{\theta_\tau}(\mathbf{x}_{i_{\tau}})\rangle > 1 \\[5pt]
        (1 - \frac{1}{\tau })\mathbf{w}_{\tau-1} + \frac{1}{\lambda\tau}y_{i_{\tau}}\psi_{\theta_\tau}(\mathbf{x}_{i_{\tau}}), & \textnormal{otherwise}.
    \end{cases} \\
    &= \frac{1}{\lambda\tau}\sum_{t=1}^{\tau}\alpha_ty_{i_t}\psi_{\theta_t}(\mathbf{x}_{i_t})\, ,
\end{align*}
where $\alpha_{\tau} = \mathbbm{1}\left[y_{i_{\tau}}\langle\mathbf{w},\psi_{\theta_\tau}(\mathbf{x}_{i_{\tau}})\rangle \leq 1\right]$ for $\mathbbm{1}[\cdot]$ being the indicator function. Having found the value of $\alpha_{\tau}$, we determine the trainable parameters for the next step $\theta_{\tau + 1}$. Note that if we are in the regime where the hinge-loss in~\cref{eq:partial_obj} is 0, the gradient with respect to $\theta$ will vanish. In the non-zero hing-loss regime, we can approximate the gradient in the $\theta$-direction numerically, e.g.~with  simultaneous perturbation stochastic approximation (SPSA). A detailed instruction of the algorithm is provided in~\Cref{algo:pegasos_qka}.

\begin{algorithm} 
	\caption{Kernel Alignment with \Pegasos} 
	\label{algo:pegasos_qka}
	\begin{algorithmic}[1]
		\STATE \textbf{Inputs:}
		\STATE training data $T = \{\mathbf{x}_1, \mathbf{x}_2, ..., \mathbf{x}_M\}$
		\STATE labels $L = \{y_1, y_2, ..., y_M\}$
		\STATE regularization parameter $\lambda \in \R^+$
		\STATE pseudo-Kernel of the form $k_{\theta_1,\theta_2}(\mathbf{x},\mathbf{y}) = \left|\braket{\psi_{\theta_2}(\mathbf{x})}{\psi_{\theta_1}(\mathbf{y})}\right|^2$
		\STATE initial kernel parameters $\theta_1 \in \R^d$
		\STATE number of steps $\tau \in \N$
		\STATE number of initialization steps $\tau_{in} < \tau$
		\STATE
		
		\FOR{$t = 1, 2, ..., \tau$}
		\STATE Choose $i_t \in \{1, ..., M\}$ uniformly at random.
    \IF{$t=1$}
    \STATE $\alpha_t \gets 1$
    \STATE$\theta_{t+1} \gets \theta_{t}$ 
    \ELSE
		
		\IF{$y_{i_{t}} \frac{1}{\lambda t} \sum_{s=1}^{t-1} \alpha_{s} y_{i_s} k_{\theta_s,\theta_t}\left(\mathbf{x}_{i_{s}}, \mathbf{x}_{i_t}\right)<1$} \label{pegasos:condition}
		\STATE $\alpha_t \gets 1$
		\IF{$t > \tau_{in}$}
		\STATE $f(\theta) \gets -y_{i_{t}} \frac{1}{\lambda t} \sum_{s=1}^{t-1} \alpha_{s} y_{i_s} k_{\theta_s,\theta}\left(\mathbf{x}_{i_{s}}, \mathbf{x}_{i_t}\right)$
		\STATE $\theta_{t+1} \gets$ Peform minimization step on $f(\theta)$ around $\theta_t$, e.g. using SPSA.
		\ELSE
		\STATE$\theta_{t+1} \gets \theta_{t}$ 
		\ENDIF
		
		\ELSE
		\STATE $\alpha_t \gets 0$
		\STATE $\theta_{t+1} \gets \theta_{t}$ 
		\ENDIF
  \ENDIF

		\ENDFOR
	\end{algorithmic}
\end{algorithm}
\subsection{Kernel alignment for non-stationary data}
\label{sec:qka_non_stationary}
The formulation in~\cref{eq:pegasos_it_sum} provides a clear chronological structure to how the weights $\mathbf{w}$ are defined. This is useful in online machine learning, where the structure of the data is time dependent and training data becomes outdated after a certain period. In standard machine learning models, it is usually easy to add new training data and continue training an existing model. However, it is often unclear how to remove or unlearn the impact of outdated training data. For that reason, models are usually completely retrained after a some time in order to adjust to the change in the data structure. This is not the case for classification with the \Pegasos~algorithm: If we realise that the structure in our training data has changed significantly, we can simply discard past iterations (and the corresponding data points) and only sum over the relevant time period by setting $\alpha_t = 0$ for $t$ lying far in the past.
\section{Experimental demonstration} \label{sec_experiments}
In this section we demonstrate that the proposed algorithm succeeds in solving the kernel alignment task proposed in \cite{Glick2021}. For a detailed derivation and motivation of the specific classification problem, we thus refer to \cite{Glick2021}. Crucially, given a device with $n$ qubits and connectivity graph $G(V,E)$, a data set $\mathcal{D}(n,G)$ with $n$ features $\mathbf{x} \in \R^n$ and binary labels $y \in \{-1,1\}$ is created. We classify $\mathcal{D}(n,G)$ using the feature map $\ket{\psi_\theta(\mathbf{x})} = U(\mathbf{x})V_\theta \ket{0}$, where $U(\mathbf{x}) = \bigotimes_{k=1}^n R_X(x_{2k-1})R_Z(x_{2k})$ and $V_\theta = \big(\prod_{(kl) \in E} \text{CNOT}(k,l)\big)\big(\bigotimes_{k=1}^n R_Y(\theta)\big)$. For the ideal parameter $\theta = \theta_{\text{opt}} = \pi/2$ the data set becomes linearly separable and the QSVM can thus reach 100\% accuracy, however if $\theta$ is chosen badly, the data cannot be separated, resulting in a lower training accuracy.

\subsection{Stationary data}
\label{sec:experiments_stationary}
\begin{figure}[!htb]
    \centering
    \includegraphics[width=\linewidth]{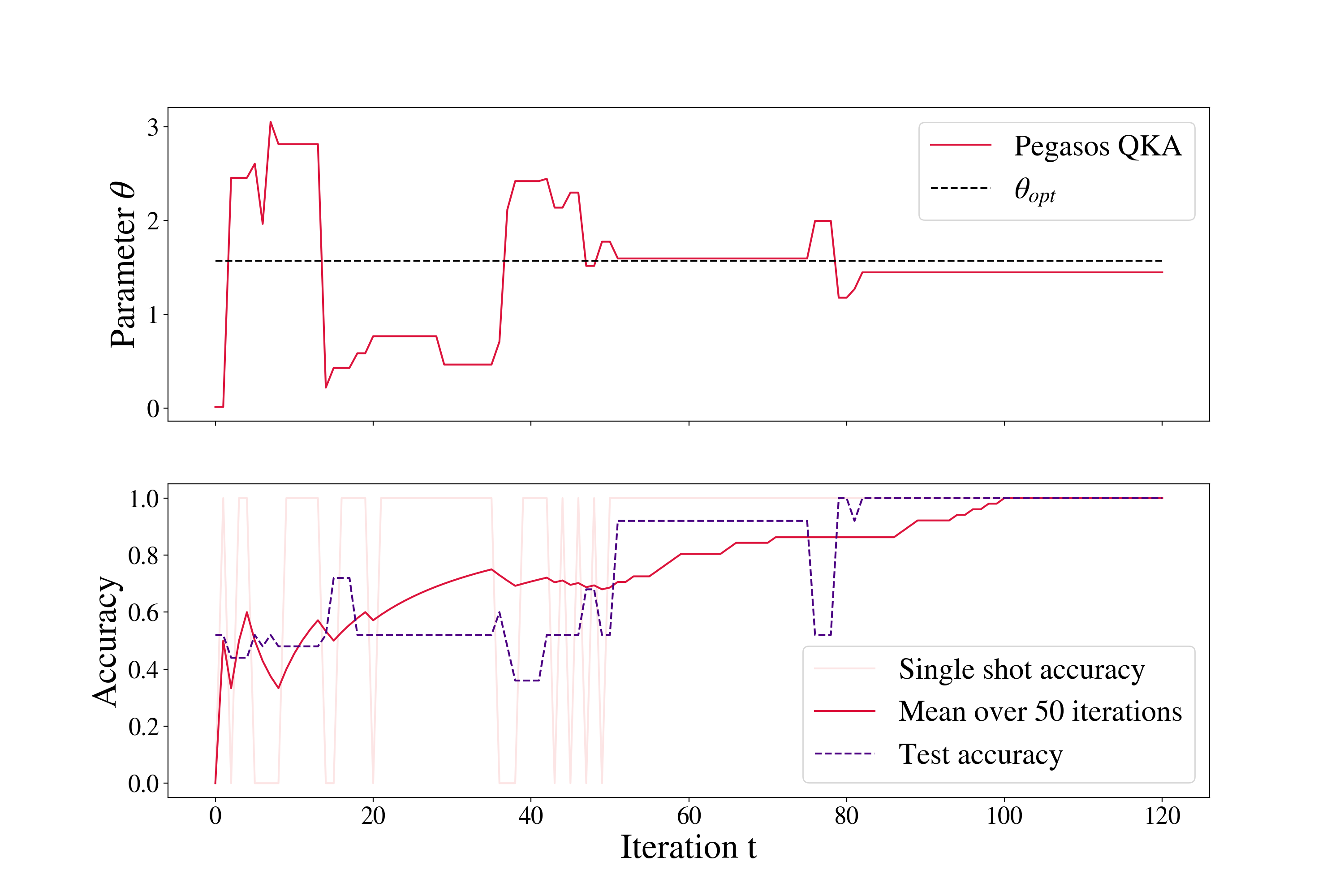}
    \caption{\textbf{Quantum kernel alignment with \Pegasos} using statevector simulation on 10 qubits using Qiskit. We train \Pegasos ~on the data provided in \cite{Glick2021} to classify the training data correctly while aligning the kernel. The top plot shows how the trainable parameter is trained using SPSA with a learning rate of $\mu=0.1$. The bottom plot shows if the sampled data points are classified correctly for every iteration (single shot accuracy) and averaged over the last 50 iterations, as well as the test accuracy on 25 unseen data points.}
    \label{fig:pegasosqka}
\end{figure}

In this fist experiment, we train \Pegasos\ on $\mathcal{D}(n,G)$. We initialize $\theta = 0$ and perform the simultaneous optimization of $\theta$ and $\boldsymbol{\alpha}$ as described in~\Cref{sec_QKA_Pega}. \Cref{fig:pegasosqka} shows that we are able to reach 100\% accuracy.

\subsection{Non-stationary data}
\label{sec:experiments_non_stationary}
\begin{figure}[!htb]
    \centering
    \includegraphics[width=\linewidth]{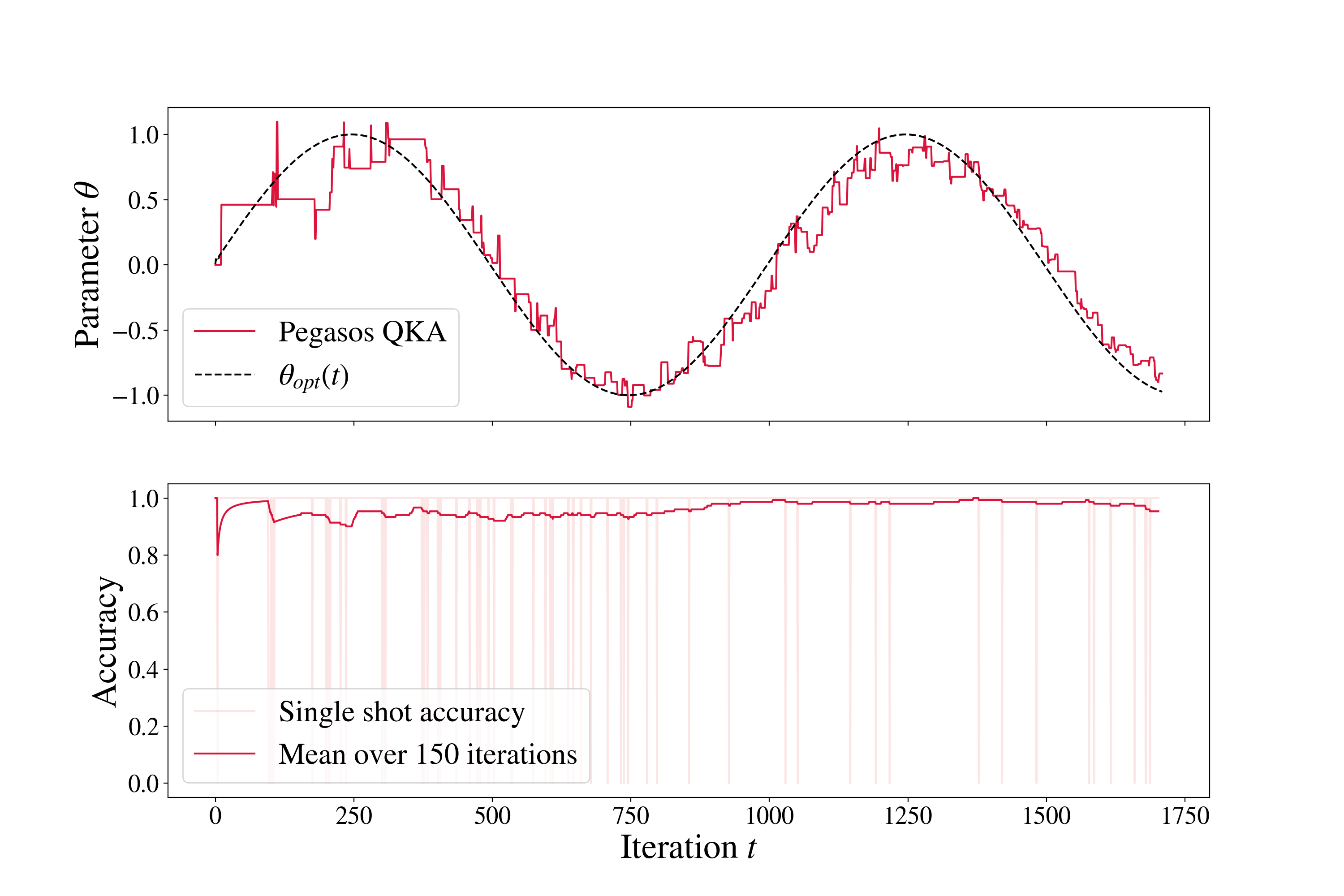}
    \caption{\textbf{Quantum kernel alignment with \Pegasos\ for non-stationary data}. In every iteration we modify $\theta_\text{opt}$ defining the structure in the data and draw a random data point from the corresponding training set as defined in \cite{Glick2021}. We let $\theta_\text{opt}$ change continuously according to a sine wave and see how \Pegasos QKA is able to track this parameter, while keeping up the accuracy of the classification. Experiment on 7-qubit device \texttt{ibm\_nairobi} with SPSA learning rate $\mu=0.1$ and training window $\tau=100$.}
    \label{fig:drifting_lambda_7}
\end{figure}

In a second experiment, we test the algorithm on non-stationary data. For this, we slightly change the setup from the above experiment by changing the structure of the data set as a function of time. Instead of fixing the parameter $\theta_{\text{opt}} = \pi/2$ used to generate the data, we allow $\theta_{\text{opt}}$ to vary every time a new tuple $(\mathbf{x_i}, y_i)$ is sampled like
\begin{equation*}
    \theta_{\text{opt}}(t) = \sin(2\pi t/T)\, ,
\end{equation*}
where $t$ indicates the current iteration count and $T = 1000$. As described in \Cref{sec:qka_non_stationary}, we only keep a window of $\tau$ iteration steps in memory in order to forget the impact of outdated training data. The goal is to  correctly classify new samples while keeping the structure parameter $\theta$ close to $\theta_{\text{opt}}(t)$. \Cref{fig:drifting_lambda_7} shows a hardware experiment demonstrating that \Pegasos\ is able to track the change in the data structure by continuously tracking $\theta_{\text{opt}}$ and classifying the newly sampled points correctly with an accuracy over 90\%.

\section{Conclusion}

Quantum kernel alignment is a promising technique for fine-tuning quantum kernels and the Pegasos algorithm allows for an elegant implementation which simultaneously performs kernel alignment as the primal QSVM problem is being solved. The speedup obtained over embedding one optimization loop within another and extends the applicability of kernel alignment to larger QSVM problems. In addition, Pegasos naturally supports online machine learning and non-stationary data as new data can easily be added to the training set and old data can be easily unlearned at any point. Our experiments demonstrate that Pegasos with simultaneous kernel alignment works well in practice and we expect this technique to be of value as QSVM models are scaled to larger and more realistic problems. 

\paragraph{Code availability}
The code for our experiments presented in~\Cref{sec_experiments} has been written using Qiskit~\cite{qiskit} and is available at~\cite{Code}.
\paragraph{Acknowledgements}
We thank Dave Clader and Rajiv Krishnakumar for helpful discussions on quantum kernel alignment and in particular to the relevance on non-stationary data in practice. For technical support with the simulations and hardware experiments we thank Elena Peña Tapia and Julien Gacon. G.G. acknowledges support by the NCCR MARVEL, a National Centre of Competence in Research, funded by the Swiss National Science Foundation (grant number 205602).

\bibliographystyle{arxiv_no_month}
\bibliography{notes}

\end{document}